\documentclass[preprint]{aastex}
\usepackage{lscape} 
\usepackage{amsmath, amsthm,amssymb}
\usepackage{epstopdf}

\newcommand{\dense}{\ensuremath{\mbox{ g cm$^{-3}$}}}

\title{Quaoar: a Rock in the Kuiper Belt}

\author{Wesley C. Fraser {1},
Michael E. Brown {1},
}
\altaffiltext{1}{Division of Geological and Planetary Sciences, California Institute of Technology, MC 150-21. 1200 E. California Blvd. Pasadena, CA 91125.}
\date{} 

\accepted{March 30, 2010}

\begin{abstract}
Here we report WFPC2 observations of the Quaoar-Weywot Kuiper belt binary. From these observations we find that Weywot is on an elliptical orbit with eccentricity of $0.14 \pm 0.04$, period of $12.438\pm0.005$ days, and a semi-major axis of $1.45\pm0.08 \times 10^4$ km. The orbit reveals a surpsingly high Quaoar-Weywot system mass of $1.6\pm0.3 \times 10^{21}$ kg. Using the surface properties of the Uranian and Neptunian satellites as a proxy for Quaoar's surface, we reanalyze the size estimate from Brown and Trujillo (2004). We find, from a mean of available published size estimates, a diameter for Quaoar of $890\pm70$ km. We find Quaoar's density to be $\rho = 4.2\pm1.3 \dense$, possibly the highest density in the Kuiper belt.
\end{abstract}

\begin{document}

\maketitle

\textbf{Subject Keywords:}  Kuiper belt Objects: Individual Quaoar, general

\section{Introduction \label{sec:intro}}

Kuiper belt objects are icy members of a planetesimal population of our Solar system beyond Neptune. To date, these bodies have appeared to fall mainly into two density classes: the small KBOs have low densities, $\rho \lesssim 1 \dense$, suggesting that these objects are largely composed of ices \citep{Margot2004,Stansberry2006,Grundy2007,Grundy2008}. The largest objects measured have densities $\rho \gtrsim 2 \dense$ \citep{Buie1997,Brown2006} with the highest observed density, that of Haumea, being $\rho~2.6 \dense$  \citep{Rabinowitz2006,Lacerda2007}  suggesting a higher rock content, but still a moderate fraction of ice in the body. These two density classes suggest separate formation path-ways resulting in high densities for the largest objects, and low densities for smaller objects.

The discovery of Quaoar's satellite, Weywot \citep{Brown2007IAUC} provides an opportunity to determine the mass and density of a Kuiper belt object smaller than the large, high density objects, but larger than the small objects which have low densities. To that end, the goal of this work was to determine Weywot's orbit, and Quaoar's mass and density. Here we report observations of Quaoar using the Wide-Field Planetary Camera 2 (WFPC2) aboard the Hubble Space Telescope. In Section~\ref{sec:Observations} we report the observations and procedure for measuring Weywot's orbit. In Section~\ref{sec:size} we present a re-analysis of past measurements of Quaoar's size, and in Section~\ref{sec:discussion} we discuss possible formation scenarios for this peculiar Kuiper belt binary.

\section{Observations and Analysis \label{sec:Observations}}
\subsection{Data Reductions}
HST images were  taken on 7 separate epochs, the discovery epoch in cycle 14 ( HST GO Program 10545), one in cycle 15 (HST GO Program 10860) and 5 in cycle 16 (HST GO Program 11169). The cycle 14 observations consisted of two 300 s exposures with ACS in the optical F606w broadband filter. The cycle 15 observations consisted of 8 400 s exposures with WFPC2 in the F606w broadband filter. For the cycle 16 exposures,  in each epoch, 4 images were taken, 2 in each of the F606w and F814w filters, with 400, and 500 s exposures respectively, all with WFPC2. The multiple exposures allowed visual identification and removal of cosmic rays. The standard reduced images provided by the HST reductions pipeline were used for our analysis.

PSF fitting and subtraction were used to reveal Weywot within Quaoar's image and to get accurate relative astrometry of the binary members. Tinytim PSFs \citep{Krist1993} were convolved with a variable 2-D Gaussian kernel to account for image smearing effects such as charge-diffusion and telescope jitter and fit to the data on an image by image basis. While the PSF subtraction was able to remove the wings of Quaoar's image, the core $\sim 6$ pixel region was not properly removed; subtraction residuals were too high in amplitude to reveal the faint satellite if it fell within this region. 

\begin{figure}[p] 
   \centering
   \includegraphics[width=7in]{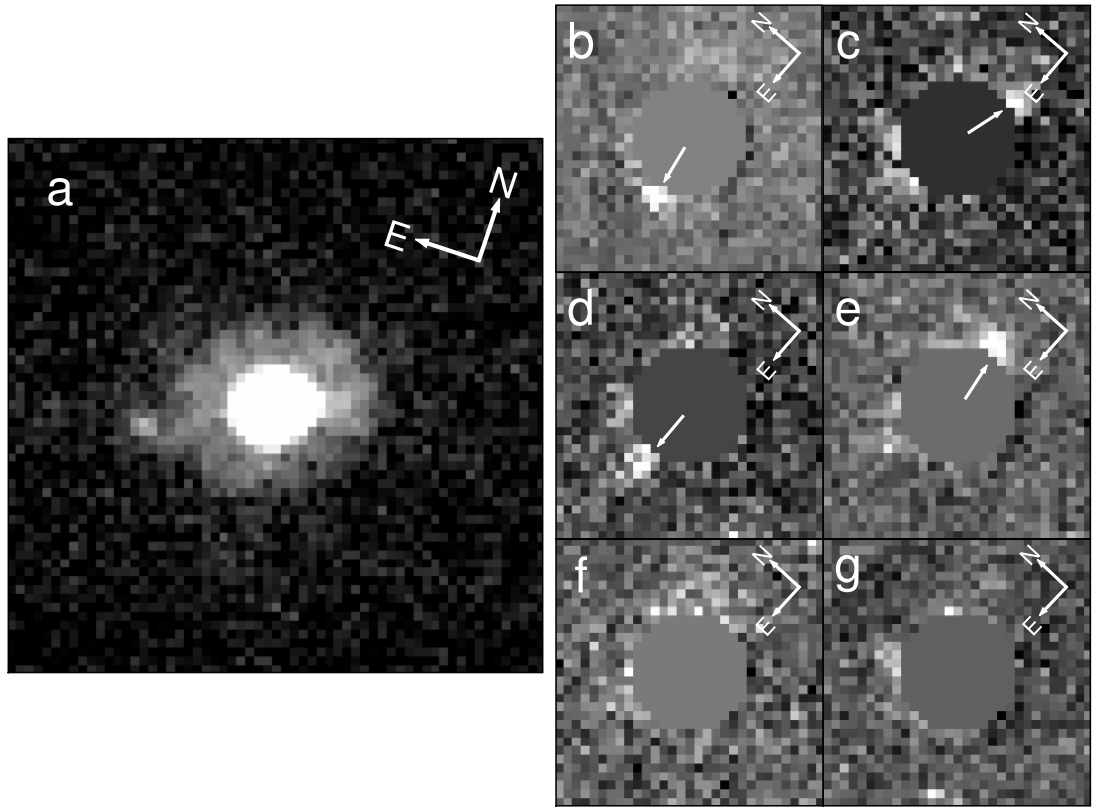}
   \figcaption{Images of the Satwot system. Letters correspond to the positions listed in Table~\ref{tab:centroids}. Orientation of the images is shown. \textbf{a} - ACS discovery image. Weywot is visible directly above the bright image of Quaoar. \textbf{b-f} - median combined WFPC2 images at each epoch.The central region in which the subtraction residuals are apparent has been masked to make the image of Weywot apparent. Weywot (marked with arrows in the WFPC2 images) can be seen as a spatially consistent point source near the top-left and bottom-right edges of the masked regions in images \textbf{b-e}. Epoch \textbf{a} has 8 images in F606w and none in F814w, while all other epochs have 2 in each filter.  \label{fig:images}}
\end{figure}

Weywot was visually identified by its common position with respect to Quaoar in all images of a common epoch. Weywot was found in the cycle 15 observations, as well as 3 out of the 5 epochs of the cycle 16 observations (see Figure~\ref{fig:images}). At each epoch, no motion of Weywot relative to Quaoar was detected. We present average positions at each epoch with uncertainties taken as the scatter in Weywot's measured position in the images at a common epoch in Table~\ref{tab:centroids}. 

From photometry derived from the PSF fitting, Quaoar has an average F606w magnitude and F606w-F814w colour of $18.96\pm0.01$ and $0.94\pm 0.02$ in the WFPC2 flight system \citep{Dolphin2000}. A variation of 0.17 magnitudes in Quaoar's brightness was observed, consistent with the peak-to-peak amplitude of the light-curve observed by \citet{Ortiz2003}. No significant variation in the colour was observed. From the original ACS discovery images, the apparent fractional brightness of Weywot compared to Quaoar was 0.6\% \citep{Brown2008}. From the WFPC2 images, Weywot was found to be roughly 5 magnitudes fainter than Quaoar, in agreement with the ACS images. This implies a size ratio approximately 12:1 and a mass ratio of $\sim$2000:1 assuming equal albedos and densities. Due to its faintness and proximity to Quaoar, photometry of Weywot of a satisfactory quality was not possible from the WFPC2 observations. 

\subsection{Orbit Fitting}
Including the 2006 discovery epoch, and the observations presented here, 5 detections and two non-detections are available from which to fit Weywot's orbit. Weywot was not detected in the 2002 ACS-HRC UV observations of Quaoar \citep{Brown2004b}. But because Weywot's UV flux is uncertain, we do not know if the observations were simply not sensitive enough to detect Weywot, or if Weywot was too close to Quaoar to be detected at that epoch. Therefore, the ACS UV observations could not be used in determining Weywot's orbit.

For the orbit fitting procedure, we adopted a Bayesian likelihood technique. The probability of  detection of the satellite at a certain position was treated as a Gaussian distribution, with mean position evaluated from the orbit at the epoch of detection, and a standard deviation equal to the uncertainty in the detection centroid. It was assumed that for the epochs in which Weywot was not detected, the satellite image fell within Quaoar's poorly subtracted image core. We chose priors that reflected this; we set a probability of 1 for orbits that placed the satellite within 0.3" or 6 pixels of Quaoar's centroid during the non-detection epochs, and a probability of 0 if the satellite was sufficiently far from the core that it could be detected given the data. We note here that, even though the best-fit orbits naturally placed Weywot within Quaoar's core during the non-detection epochs\footnote{If the non-detections are entirely ignored in the fitting process, the orbit and resultant system mass are identical to the nominal orbits we present here.}, other effects could have caused those non-detections, such as if Weywot posses a large amplitude light curve. The necessary dip in
 flux for a non-detection however, is about a factor of 10. Therefore we discount this possibility.

The observations were fitted with circular and elliptical orbits. We determined if the fits were satisfactory using a maximum likelihood distribution test. That is, using the best-fit orbits, random positions of Weywot were generated with identical temporal spacing as in the observations, and were refit using the same maximum likelihood technique as used on the original data. This procedure was repeated 1000 times and the distribution of likelihood values was recorded. If the best-fit is an acceptable description of the data, then the distribution of simulated likelihood values would bracket the real value, with less than 67\% (1-standard deviation) of the simulated values larger than the actual value.

It was found that for the best-fit circular orbit, the probability of finding a simulated maximum likelihood value greater than that from the observations was 99.7\%, which is a 3-$\sigma$ deviation from the expected range of likelihoods. This result demonstrates that the best-fit circular orbit is an unsatisfactory fit to the observations, and is rejected at the 3-$\sigma$ confidence level. Rather, Weywot must be on an elliptical orbit. The probability of finding a simulated maximum likelihood value larger than that for the elliptical orbit was 10\%, indicating that the elliptical orbit is a satisfactory description of the observations. The low probability however, suggests that the astrometric errors adopted here are slightly too large, but not large enough to warrant reanalysis.

\begin{figure}[h] 
   \centering
   \plotone{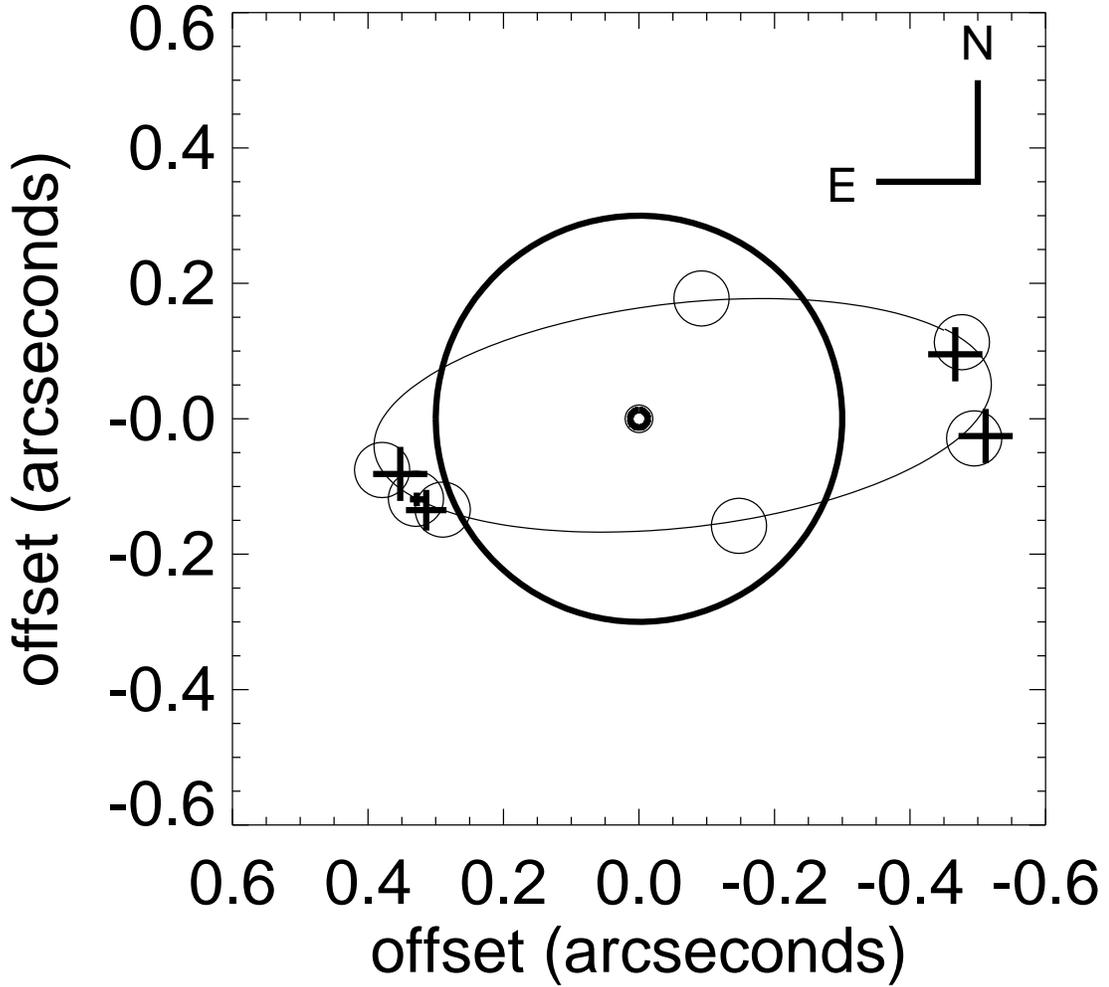} 
   \figcaption{Best-fit orbit and offsets of Weywot from Quaoar's image center. Orbital parameters are presented in Table~\ref{tab:orbits}. Crosses along the ellipse represent the measurement and uncertainty of Weywot's position at each epoch (see Table~\ref{tab:centroids}). The small cross is the satellite discovery detection \citep{Brown2007IAUC}. The large thick-lined circle represents the core region in which the satellite would not have been detected. The small circles mark the locations of the satellite along the inferred orbit at the observation epochs.   \label{fig:orbit}}
\end{figure}

Due to the projection of Weywot's orbit on the sky, two equally satisfactory fits to the observations can be found. This degeneracy does not affect the system mass determination however, as both orbits have virtually identical periods and semi-major axes. The best-fit orbits are presented in Figure~\ref{fig:orbit}, with parameters presented in Table~\ref{tab:orbits}. The orbits have a period of $12.438\pm0.005$  days and a semi-major axis of $1.45 \pm 0.08 \times 10^4$ km and imply that the Quaoar-Weywot system mass is $1.6\pm0.3 \times 10^{21}$ kg, or roughly 12\% that of Eris \citep{Brown2007b}.  

\section{Quaoar's Size \label{sec:size}}

Quaoar's size has been measured by two different methods. Partially resolved images of Quaoar from the Hubble Space Telescope suggest a diameter, $D=1260\pm190 \mbox{ km}$ \citep{Brown2004b}, while two independent size determinations  from its thermal emission observed by the Spitzer telescope give diameters $D=844_{-190}^{+207}$ \citep{Stansberry2008} and $D=908_{-118}^{+112} \mbox{ km}$ \citep{Brucker2009}.

In the case of the HST measurement, the inferred size depends on Quaoar's unknown limb-darkening function and depends roughly linearly on the half-light diameter of this function - the diameter of an aperture which contains half of the light reflected from Quaoar towards the observer \citep{Brown2004b}. At the time of the HST measurement, little was known about the surface properties of Kuiper belt objects, and a Lambert sphere was adopted for the limb darkening profile, with uncertainties reflecting the lack of knowledge of Quaoar's true limb darkening profile. The uncertainties in the HST measurement are dominated by this unknown.

We now know that the surface of Quaoar appears in many ways similar to those of the icy satellites of Uranus and Neptune. They exhibit similar ice-absorption features \citep{Cruikshank1986,Schaller2007}, have high albedos \citep{Karkoschka2001,Brown2004b}, and have similar optical colours. Quaoar's opposition surge, with slope $\beta(V)=0.16\pm0.03 \mbox{ mag deg$^{-1}$}$ is also similar to that of the Uranian satellites, with slopes in the range $\beta(V)=0.07-0.18 \mbox{ mag deg$^{-1}$}$ \citep{Karkoschka2001, Rabinowitz2007,Belskaya2008}. Adopting a Uranian-satellite limb darkening profile rather than a Lambert profile appears to be a better approach. From the fits of the Hapke surface reflectance model \citep{Hapke2002} to the icy satellite phase functions \citep{Karkoschka2001}, we have calculated the limb-darkening profiles of the satellites and found their half-light diameters to be between $ 0.82 d$ (that of Triton) and $0.88 d$ (that of Umbriel), where $d$ is the diameter of the object in question. The adopted half-light diameter - that of a Lambert sphere - is $0.6 d$. This suggests that the size estimate of Quaoar inferred from the HST data  is $\sim 40\%$ too large. Adopting the average limb-darkening profile of the satellites, the Hubble observations suggest that Quaoar's diameter is $D=900$ km. While the icy satellites exhibit a wide diversity of surface properties, they only exhibit  a small range in half-light diameters, $0.82 \mbox{-} 0.88 d$, or 6\% of their mean value. Indeed, icy bodies throughout the Solar system for which sufficient data are available - Enceladus, Io, Europa, Ganymede, Callisto, and the Uranian and Neptunian satellites discussed here - span only a 9\% range \citep{Verbiscer2005, Domingue1997, Domingue1998}. Therefore, we adopt 6\% as the systematic uncertainty in Quaoar's size measurement due to our assumption about its unknown half-light diameter. Adding in quadrature this uncertainty, and the 4\% random and remaining 10\% systematic uncertainties in the Hubble measurement, the total uncertainty in Quaoar's diameter is 13\%, or $115$ km. Unlike the size inferred from adopting a lambert profile, the corrected size measurement is compatible with both Spitzer size measurements, lending confidence in those measurements and our use of the icy satellites as proxies for Quaoar's surface. We adopt as a size of Quaoar, the weighted average of the Spitzer and corrected HST estimates, $D_{\rm{Quaoar}} = 890 \pm 70$ km.

The adopted size and inferred mass of Quaoar imply that Quaoar's density is $\rho = 4.2\pm1.3 \dense$. This is only marginally compatible with the density of the next most dense Kuiper belt object, Haumea \citep{Rabinowitz2006,Lacerda2007} and we conclude that Quaoar is likely the most dense Kuiper belt object currently known.


\section{Discussion \label{sec:discussion}}
 Quaoar's unusually high density implies that this Kuiper belt object has little ice content. A thin veneer of surface ice is required for Quaoar to exhibit its absorption features indicative of water, methane, and ethane ices \citep{Jewitt2004,Schaller2007}. But this ice cannot be a substantial component of the body's mass.


Quaoar's high density is reminiscent of the asteroid belt. It may be possible that, in the early Solar system, during some dynamical event, such as the migration of Jupiter, a series of scattering events emplaced asteroids into the Kuiper belt region. Indeed, it has been shown that migration can emplace Kuiper belt objects in the stable asteroid belt region \citep{Levison2009}. Thus it seems possible that the reverse process could occur suggesting that Quaoar might once have been an asteroid which lost the majority of its ice content due to rapid sublimation from Solar insolation, or never had a substantial ice content, and was scattered onto its current orbit.

It is also possible that Quaoar's high density is collisionally produced. The small satellites of the largest Kuiper belt objects, which presumably formed through massive collisions, appear icy \citep{Marcialis1987,Barkume2006,Fraser2009b}, consistent with the idea that the satellites are ejected material from the  ice surface layers of large differentiated parent bodies. Presumably,  the resultant density of the remaining body is a function of the impact properties (impact angle, velocity, etc.). Given the lack of understanding of large-scale collisions, it may be possible that with a particular impact scenario, a collision can strip nearly 100\% of the surface layer of a parent body,  leaving the rocky-core that is Quaoar virtually intact.

Weywot's orbit is difficult to explain with a collisional genesis. Ejecta form disks from which a satellite might coalesce. The satellite then tidally evolves outwards on a circular orbit rather than the elliptical orbit of Weywot.

Weywot's orbit could however, be explained if another satellite of similar mass once existed about Quaoar. Dynamical interactions could allow the two satellites to scatter off one-another, emplacing Weywot on its eccentric orbit and removing the second satellite from the system. Indeed an order-of-magnitude estimate \citep{Goldreich1966} implies that the circularization time scale of Weywot's orbit is approximately the age of the Solar system, implying that, if unperturbed, once Weywot is on an eccentric orbit, it will remain that way. Given the existence of other collisionally formed Kuiper belt binaries, this formation mechanism seems plausible.

Another mechanism that might explain the properties of the Quaoar-Weywot system is a so-called hit-and-run collision, first proposed by \citet{Asphaug2006}. In such a scenario, an originally ice-rich and differentiated Quaoar has a grazing impact with a body roughly 2-3 times as massive. The result is a complete shattering and scattering of Quaoar's icy mantle, leaving its core bound, and relatively intact. During such an impact, the prevalence of three-body interactions  between the ejecta allows many ejecta pairs to become bound with mutually eccentric orbits. Such a collision could explain Quaoar's unusually high density and Weywot's eccentric orbit. This scenario requires the primordial Kuiper belt to be significantly more massive than the present for such a collision to be likely. If this scenario were prevalent in the early Kuiper belt, it would suggest that a large fraction of small Kuiper belt objects are the mineral-less, low-density icy ejected mantle fragments of large differentiated bodies. As well we should find a few large $\sim 1000$ km bodies with high $\sim 3 \dense$ densities that were originally the cores of those larger differentiated planetesimals.

\section{Acknowledgements}
We would like to thank Hal Levison, Eric Asphaug, Darin Ragozzine, and Alex Parker for their insightful discussions.

This material is based upon work supported by NASA under grant NNG05GI02G. Support for program HST-Go-011169.9-A was provided by NASA through a grant from the Space Telescope Science Institute, which is operated by the Association of Universities for Research in Astronomy, Inc., under NASA contract NAS 5-26555.


\begin{deluxetable}{ccc}
   \tablecaption{Quaoar and Satellite Positions. $\Delta$ RA and $\Delta$ DEC are the differences in position between the satellite and Quaoar. The first row is from Weywot's discovery observations. For the  two cycle 16 epochs in which the satellite was not detected, it is assumed that the satellite image fell within Quaoar's core. The positional uncertainties were adopted to reflect the scatter in the measured satellite positions at each epoch.\label{tab:centroids}}
   \startdata	
      Epoch  &\multicolumn{2}{c}{Satellite-Quaoar Offsets} \\ 
   (JD+2453000) &  $\Delta$ RA (arcsec.) & $\Delta$ DEC (arcsec.) \\
      \hline
     781.38031 & $0.328 \pm 0.01$ & $-0.119 \pm 0.01$ \\
   1179.12990 & $0.314 \pm 0.03$ & $-0.135 \pm 0.03$\\
   1535.70263 & $-0.51\pm 0.04$ & $-0.02\pm 0.04$ \\ 
   1540.56061 & $0.35\pm 0.04$ & $-0.08\pm 0.04$ \\
   1546.18353 & $-0.47\pm 0.04$ & $0.09\pm 0.04$\\
   1550.31485 & - & - \\
   1556.44075 & - & - \\

   \enddata
\end{deluxetable}

\begin{deluxetable}{lcc}
   \tablecaption{Best-fit parameters for the orbit of Weywot and inferred system mass. Uncertainties are the extrema of the 6-parameter likelihood volume which contain 67\% of the total likelihood integral. All values are relative to the J2000 ecliptic. \label{tab:orbits}}
   \startdata	
   Orbital Parameter & Orbit 1 & Orbit 2 \\
   \hline
   Period (days) & $12.438\pm0.005$ & $12.439\pm0.005$  \\
   Semimajor Axis $(10^4 \mbox{ km})$ & $1.45 \pm 0.08$ & $1.45 \pm 0.08$ \\
   Eccentricity & $0.14 \pm 0.04$ & $0.13 \pm 0.04$ \\
   Inclination (deg.) & $14\pm4 $ & $150\pm4 $ \\
   Longitude of ascending node (deg.) & $1\pm5$ & $1\pm5$\\
   Argument of Perihelion (deg.) & $349\pm7$ & $347\pm7$\\
   System Mass ($10^{21}$ kg) & $1.6\pm0.3$ & $1.6\pm0.3$ \\
   \enddata
\end{deluxetable}

\end{document}